\documentclass[10pt]{spie}  

\usepackage{amsmath}
\usepackage{amsfonts}
\usepackage{amssymb}
\usepackage{comment}
\usepackage{graphicx}
\usepackage{hyperref}
\usepackage{xcolor}
\usepackage{float}
\floatstyle{plaintop}
\restylefloat{table}

\title{Mueller Matrix Decomposition for a General Rotating Compensator Ellipsometer}
\author{Aaron Chew, Purnima P. Balakrishnan, Bryan M. Barnes, John H. Burnett, Colton K. Dudley, Steven E. Grantham,  Stephanie L. Moffitt, Charles Tarrio, Thomas A. Germer}
\affil{National Institute of Standards and Technology, 100 Bureau Dr., Gaithersburg, MD 20899 USA}

\authorinfo{Further author information: (Send correspondence to A.C.)\\A.C.: E-mail: aaron.chew@nist.gov}

\begin{document}
\maketitle

\begin{abstract}
    This work develops a Mueller matrix decomposition that factors a nondepolarizing Mueller matrix $M$ into three components $M = T_1 D T_2$, where the elements $D$, which correspond to a diagonal Jones matrix, are separated from rotations and circular diattenuation $T_1, T_2$. This decomposition is employed to calibrate a prototype reflective rotating-compensator ellipsometer designed to operate using extreme ultraviolet (EUV) radiation. This decomposition allows for the measurement of the full Mueller matrix of a sample without full knowledge of the compensators or the source polarization, while allowing for significant diattenuation and retardance for the optics.  A generalization of the decomposition is developed for depolarizing Mueller matrices, and the technique is tested on reflection data collected on an Au mirror at $266$ nm.  
\end{abstract}

\section{Introduction}

NIST has developed a dual-rotating-compensator spectroscopic Mueller-matrix ellipsometer that operates in the extreme ultraviolet (EUV) \cite{Balakrishnan2026}. This ellipsometer modulates the polarization of light incident on a sample with a rotating compensator optic, and the polarization of light scattered from the sample is modulated in the same way to analyze the change in polarization. In this way, the full Mueller matrix of a sample can be extracted. Typical retarding compensators for ultraviolet (UV)/visible/near-infrared (NIR) wavelengths have well-characterized transmissive optics.  However, because most materials are strongly absorbing in the EUV, all optics in the EUV ellipsometer must be reflective instead of transmissive. Thus, each compensator is a series of mirrors arranged as a chevron \cite{Moffitt2024, Chew2025}, as depicted in Fig.~\ref{fig:ellipsometer}. Unlike typical retarding compensators for UV/visible/NIR wavelengths, each chevron acts as a combination retarder and diattenuator; this behavior is described by a Mueller matrix $M_\text{chevron}$, which depends heavily on reflection angle, wavelength, and mirror surface \cite{Chew2025, Tarrio2023}.  Our chevrons in general possess significant diattenuation and retardance.  Imperfections in mirror alignment as well as surface changes, such as carbon growth after exposure to EUV radiation, make $M_\text{chevron}$ not only unknown, but also potentially unstable over long time scales. This necessitates a measurement scheme which can be employed to calibrate the system without prior knowledge.

Light with initial Stokes polarization $Q$ is modulated by rotating the chevron about its optical axis, resulting in light with polarization $W$ incident on a sample. In the Stokes-Mueller formalism, this rotation is expressed as
\begin{align}
W(\theta) = R(\theta) M_\text{chevron} R(-\theta) Q,
\end{align}
where $R(\theta)$ is the rotation matrix for angle $\theta$ about the optical axis.
If $W(\theta)$ is known or measured at a number of different rotations, we seek to determine $M_\text{chevron}$ and $Q$. Unfortunately, with only knowledge of $W(\theta)$, $M_\text{chevron}$ and $Q$ can never be fully identified. Consider the transformation 
\begin{align}
M_\text{chevron}' = M_\text{chevron}T,~~Q' = T^{-1} Q,
\end{align} where the matrix $T$ commutes with all rotations $R(\theta)$. The output Stokes vector $W(\theta)$ remains invariant under such a transformation, and so $(M_\text{chevron}, Q)$ cannot be distinguished from $(M_\text{chevron}', Q')$.  

To address this issue, we develop a Mueller matrix decomposition that separates its diagonal components from rotation and circular diattenuation.  Here, `diagonal' refers to reflections that preserve $s$ and $p$ polarization: in the Jones formalism, their $2 \times 2$ reflection matrices are diagonal.  This decomposition identifies which degrees of freedom of the Mueller matrix ellipsometer can and cannot be identified from a measurement, and elucidates a calibration protocol which allows the full Mueller matrix of a sample to be identified.  The decomposition is further extended to model depolarizing effects.  Our decomposition joins a large family of other Mueller matrix decompositions \cite{Azzam:78, Sridhar01101994, Lu:96, Devlaminck:08, Simon:10, Ossikovski:08, Ossikovski:09, Ossikovski:10, Vizet:18, GAO2021106735, Xing1992}, as well as techniques aimed at studying rotating compensator ellipsometers \cite{Hauge:78, KANG2026114940, BIAN2022139581, Tong:25}.   Unlike previous works, we must allow for general Mueller matrices for the chevrons, as they possess both significant diattenuation and retardance.  The versatility of this decomposition is highlighted in Section~\ref{sec:data}, where the ellipsometric quantities N, C, and S of an Au mirror are extracted without fully calibrating the ellipsometer.  The calibration technique only requires a single reflective mirror as a calibration artifact, and no polarizer, which avoids the main obstacle preventing us from employing the eigenvalue calibration technique of Ref.~\cite{Compain:99}.  We note that the form of the decomposition resembles (but is not) the retarder decomposition of Ref.~\cite{Ossikovski:08}, in which a diagonal Mueller matrix is sandwiched between two retarders.  To the best of our knowledge, our decompositions, both depolarizing and nondepolarizing, have not yet been investigated in the literature.  

\section{Matrix Decompositions}
This section summarizes our two matrix decompositions: one for nondepolarizing components (section \ref{sec:nondepolarizing}), and one which allows for depolarization (section \ref{sec:depolarizing}). Detailed proofs of their construction are given in Appendices~\ref{app:nondepolarizing} and \ref{app:depolarizing}.  Both factor a Mueller matrix into the form $T_1DT_2$, where $T$ commutes with rotations and $D$ does not.
Take a Stokes vector 
\begin{align}
\begin{bmatrix}
    I & Q & U & V
\end{bmatrix}^T.
\end{align}
The rotation matrix $R(\theta)$ is defined to be
\begin{align}
R(\theta) = \begin{bmatrix}
    1 & 0 & 0 & 0 \\
    0 & \cos 2\theta & -\sin 2\theta & 0 \\
    0 & \sin 2\theta & \cos 2\theta & 0 \\
    0 & 0 & 0 & 1
\end{bmatrix}
\end{align}
where $\theta$ is the angle of rotation about the beam propagation path.

\subsection{Nondepolarizing Matrices}
\label{sec:nondepolarizing}

Define the generalized rotation matrix $T$ (a combination of circular diattenuation and rotation) as 
\begin{align}
T(\alpha, \beta) &= \begin{bmatrix}
\cosh 2\alpha & 0 & 0 & \sinh 2\alpha \\
0 & \cos 2\beta & -\sin 2\beta & 0 \\
0 & \sin 2\beta & \cos 2\beta & 0 \\
\sinh 2\alpha & 0 & 0 & \cosh 2\alpha
\end{bmatrix}.
\label{eq:rotboost}
\end{align}

These $T$ matrices can be understood as the boost and rotation aspect of the Lorentz group \cite{Devlaminck:08, Simon:10}, and can be identified as circular diattenuation and rotations \cite{app11156742}.  While the $\beta$ components in $T$ are clearly a rotation, the $\alpha$ components correspond to circular diattenuation.  As expressed in Eq.~\ref{eq:rotboost}, the $T$ matrix is not physical as it has optical gain ($\cosh 2\alpha \geq 1$); a normalization factor of $e^{-2|\alpha|}$ may be multiplied to the matrix if one wishes to use physically realizable Mueller matrices \footnote{In the calibration procedure, it is not a concern.  All multiplicative constants are gathered into one prefactor, which is fit against data.}. All elements of this group satisfy
\begin{align}
\left[T(\alpha, \beta), R(\theta)\right] &= 0 \text{~for all~} \theta\\
T(\alpha_1, \beta_1) T(\alpha_2, \beta_2) &= T(\alpha_1 + \alpha_2, \beta_1 + \beta_2),
\end{align} where the brackets in the first line denote a commutator.  For a nondepolarizing matrix $M_\text{nondepol}$, i.e., one fully described by a complex $2 \times 2$ Jones matrix, we claim there is a decomposition 
\begin{align}
    M_\text{nondepol} = T(\alpha_1, \beta_1)~M_\text{diag}(\tau, N, C, S)~T(\alpha_2, \beta_2) \label{eq:nondepol}
\end{align} such that the diagonal Mueller matrix $M_\text{diag}$ reads
\begin{align}
M_\text{diag}(\tau, N, C, S) &= \tau \begin{bmatrix}
    1 & N & 0 & 0 \\
    N & 1 & 0 & 0 \\
    0 & 0 & C & S \\
    0 & 0 & -S & C
\end{bmatrix}, ~~~~N^2 + C^2 + S^2 = 1.
\label{eq:diagonal}
\end{align}
Thus, $M_\text{nondepol}$ is fully described by 7 independent parameters. The derivation of this decomposition is given in Appendix~\ref{app:nondepolarizing}.

\subsection{General Depolarizing Matrices}
\label{sec:depolarizing}
The above transformation can be further generalized if the optical elements are depolarizing. In this case, the circular diattenuation is no longer characterized by a single parameter $\alpha$, and the $4 \times 4$ matrices that commute with $R(\theta)$ instead read
\begin{align}
L(U, \theta) = \begin{bmatrix}
    u_{11} & 0 & 0 & u_{12}\\
    0 & ~\cos 2\theta & -\sin 2\theta & 0\\
    0 & \sin 2\theta & \cos 2\theta & 0 \\
    u_{21} & 0 & 0 & u_{22} \\
\end{bmatrix}, ~~~ U = \begin{bmatrix} 
    u_{11} & u_{12} \\
    u_{21} & u_{22} 
\end{bmatrix}.
\end{align}  These matrices $L$ obey 
\begin{align}
\left[L(U, \theta'), R(\theta)\right] &= 0 ~~\text{for all}~~\theta \\
L(U, \theta_1) L(V, \theta_2) &= L(UV, \theta_1 + \theta_2),
\end{align}  as $L$ is block diagonal.  We claim a general Mueller matrix $M$ may be decomposed as 
\begin{align}
M = L(U, \theta_1) M_\text{special}^{(p_1, p_2)}(\tau, N, C, S, A, B) L(V, \theta_2),
\label{eq:depol}
\end{align}  where $M_\text{special}^{(p_1, p_2)}$ takes the form 
\begin{align}
M_\text{special}^{(p_1, p_2)}(\tau, N, C, S, A, B) = \tau
\begin{bmatrix}
    1 & p_1 N & 0  & B \\
    N & 1 & 0 & 0 \\
    0 & 0 & C & S \\
    A & 0 & -p_2 S  & C
\end{bmatrix},~~p_1 = \pm 1,~ p_2 = \pm 1.
\end{align} Thus, the 4$\times$4 $M$ is fully  described by 16 parameters, and the signs $p_1, p_2$ are uniquely determined by the other 16 parameters. The proof of the existence of this decomposition is given in Appendix~\ref{app:depolarizing}.  

Unlike the nondepolarizing case, there is no reason to expect that the matrices $L, M_\text{special}$ correspond to physical processes; indeed, tests of the decomposition for random Mueller matrices often yield $L, M$ that are unphysical.  Nevertheless, the decomposition in Equation~\ref{eq:depol} remains mathematically correct and serves as a way to isolate the components of the Mueller matrix that can be fit in a rotating compensator ellipsometer.

\section{Ellipsometer calibration procedure}

\begin{figure}
\centering
\includegraphics[width=0.95\columnwidth]{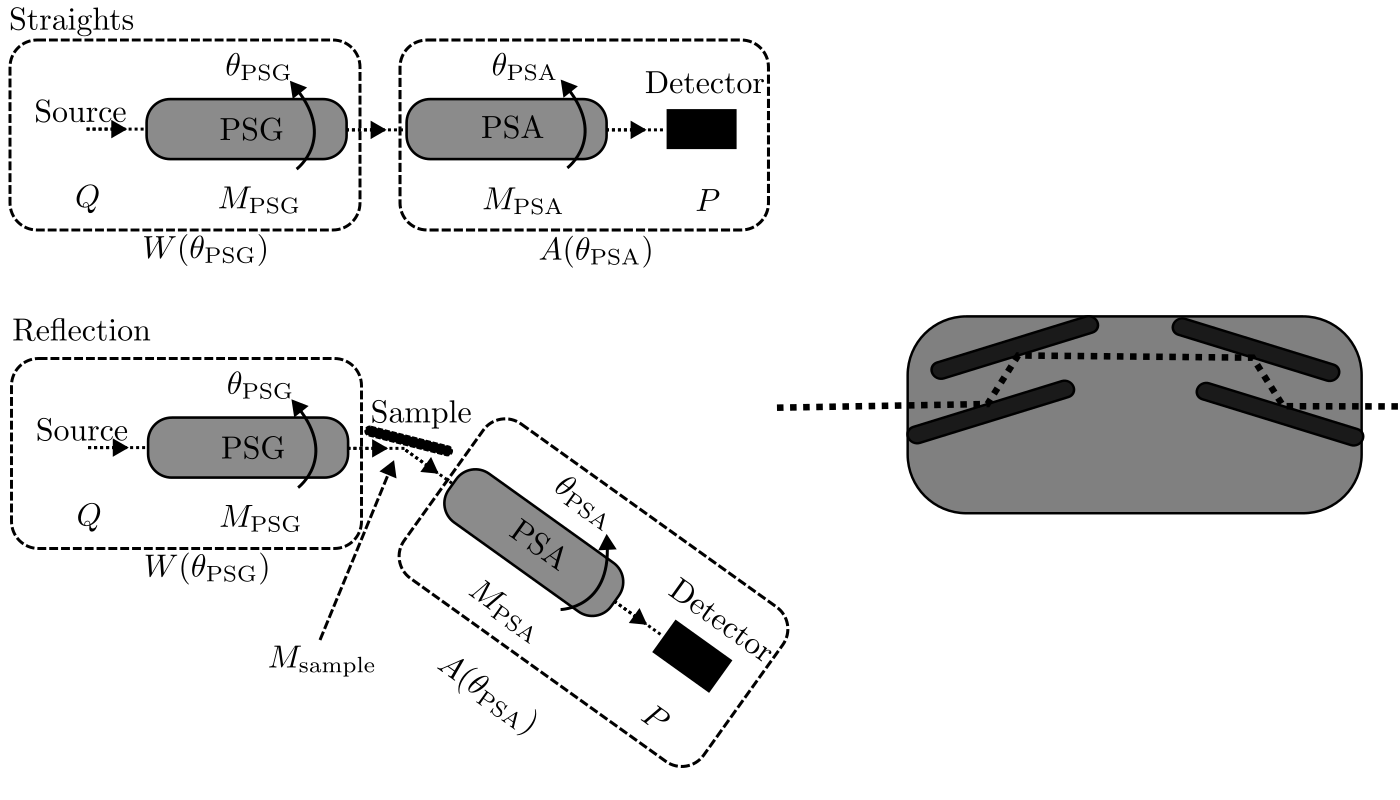}
\caption{\emph{Left:} Schematic of a rotating-compensator ellipsometer.  The optics PSA and PSG, the incoming source $Q$, and the detector $P$ must be calibrated with a fit.  The straight geometry is used to calibrate ($Q$, $M_\text{PSG}$) and ($P$, $M_\text{PSA}$), while the reflection geometry is employed to extract $M_\text{sample}$.  \emph{Right:} A four-mirror chevron geometry.}
\label{fig:ellipsometer}
\end{figure}

A simplified schematic of the dual rotating compensator ellipsometer is depicted in Fig.~\ref{fig:ellipsometer}. It consists of two rotating optics, named the polarization state generator (PSG) and polarization state analyzer (PSA). By rotating the PSA and PSG, the polarization state of the incident light $W(\theta_\text{PSG})$ and the analyzer sensitivity $A(\theta_\text{PSA})$ can be modulated; any other optics have been absorbed into the rotation-independent Stokes vectors of source polarization $Q$ and detector sensitivity $P$. If the PSG and PSA are designed appropriately, then $W(\theta_\text{PSG})$ and $A(\theta_\text{PSA})$ span enough of the Poincar\'e sphere to determine the full Mueller matrix of a sample $M_\text{sample}$.

The Mueller matrices of the PSG and PSA optics, as well as the Stokes vectors of the source and the detector, are unknown and must be calibrated.  To calibrate, a `straight' measurement is performed, in which the PSA and PSG are rotated at different angles and the output light intensity $S_\text{straight}(\theta_\text{PSA}, \theta_\text{PSG})$ is measured, without the presence of a sample:

\begin{align}
S_\text{straight}(\theta_\text{PSA}, \theta_\text{PSG}) &= A(\theta_\text{PSA}) W(\theta_\text{PSG}) \label{eq:model}\\
A(\theta_\text{PSA}) &= P R(\theta_\text{PSA}) M_\text{PSA} R(- \theta_\text{PSA}) \\
W(\theta_\text{PSG}) &= R(\theta_\text{PSG}) M_\text{PSG} R(-\theta_\text{PSG}) Q,
\end{align} where the measurement output is $S_\text{straight}$, and $P, Q$ are the Stokes vectors corresponding to the detector and incoming source, respectively.  The matrices $M_\text{PSA}, M_\text{PSG}$ are the Mueller matrices of the rotating compensators.

These straight measurements alone are not enough to fully characterize the ellipsometer -- that is, to unambiguously determine $W$ and $A$.  Decomposing the Mueller matrices of the PSA and PSG yields
\begin{align}
M_\text{PSA} = T_{\text{PSA1}} D_\text{PSA} T_{\text{PSA2}},~~M_\text{PSG} = T_{\text{PSG1}} D_\text{PSG} T_{\text{PSG2}},
\end{align} where $D_\text{PSA}, D_\text{PSG}$ have the form of Eq.~\ref{eq:diagonal}. Some of the $T$ matrices cannot be fit from the straight measurement alone:

\begin{align}
A(\theta_\text{PSA}) &= P' R(\theta_\text{PSA}) D_\text{PSA}  R(- \theta_\text{PSA}) T_{\text{PSA2}}  \\
W(\theta_\text{PSG}) &= T_{\text{PSG1}} R(\theta_\text{PSG}) D_\text{PSG}  R(-\theta_\text{PSG}) Q' \\
S_\text{straights}(\theta_\text{PSA}, \theta_\text{PSG}) &= P' R(\theta_\text{PSA}) D_\text{PSA}  R(\theta_\text{PSG}- \theta_\text{PSA}) T' D_\text{PSG}  R(-\theta_\text{PSG}) Q' \label{eq:straight}
\end{align} where 
\begin{align}
P' = P T_{\text{PSA1}}, ~~Q' = T_{\text{PSG2}} Q, ~~ T' = T_{\text{PSA2}} T_{\text{PSG1}}.
\end{align}  We have used the fact that $R$ and $T$ commute.  In other words, it is impossible to determine both $T_{\text{PSA2}}$ and $T_{\text{PSG1}}$ individually -- only their product $T'$ is known.  When fitting the straight measurement, we obtain normalized values for $P', Q', T', D_\text{PSA}, D_\text{PSG}$, and one overall multiplicative constant $\tau$.  However, when measuring a sample, one must know $T_{\text{PSG2}}$ and $T_{\text{PSA1}}$.  This is not possible with the straight measurement.

A second calibration measurement involving an object with a Mueller matrix of NCS form $M_\text{cal,diag}$ must be performed to fix $T_{\text{PSA2}}$ and $T_{\text{PSG1}}$.  The measured quantity is
\begin{align}
S_\text{cal}(\theta_\text{PSA}, \theta_\text{PSG}) =  A(\theta_\text{PSA}) M_\text{cal,diag} W(\theta_\text{PSG}).
\end{align}  So long as $M_\text{cal,diag}$ is of NCS form and not proportional to the identity (that is, it is not a basic filter that affects $s$-polarized and $p$-polarized light in the same manner), this second measurement, in addition to the straight measurement, is enough to determine $T_\text{PSA2}$ and $T_\text{PSG1}$ and thus fully calibrate a nondepolarizing ellipsometer.  An example of such an optic would be a mirror constructed of an isotropic material.  A similar requirement was noted for a calibration where the chevrons were Fresnel rhombs and were modeled as elliptical retarders by the authors of Ref.~\cite{BIAN2022139581}.

If this second calibration is \emph{not} performed, important information can still be extracted about the sample.  Assuming the nondepolarizing form for the sample, 
\begin{align}
S_\text{sample}(\theta_\text{PSA}, \theta_\text{PSG}) =  A(\theta_\text{PSA}) T_\text{sample,1} M_\text{sample,diag} T_\text{sample,2} W(\theta_\text{PSG}).
\end{align}  While $T_\text{sample,1}$ and $T_\text{sample,2}$ cannot be extracted (as $T_\text{PSA1}, T_\text{PSG2}$ are undetermined), their product can be, as well as the diagonal components $M_\text{sample,diag}$.  If one expects the sample to be close to NCS form, then $M_\text{sample, diag}$ may be extracted without knowing $T_\text{sample,1}, T_\text{sample,2}$. 

In summary, to determine the Mueller matrix for a sample that is supposed to be of NCS form (such as a thin film or multilayer), calibration with measurement in a straight geometry is sufficient. If the sample is expected to deviate from this form, then a second measurement using a diagonal calibration artifact is necessary to determine the full Mueller matrix of the sample.

\section{Generalized Matrix Decomposition}

Can general depolarizing chevrons can be calibrated according to the steps given above?  A straight measurement, allowing for depolarizing chevrons, would have $22$ floating parameters: $3$ for the input and output polarizations, $5$ for each of the chevrons ($N,C,S,A,B$), $5$ for the generalized rotation/circular dichroism matrix between the PSA and PSG, and one total normalization, for a total of $22$.  A straight measurement has $25$ different Fourier harmonics, and so naively one might expect the straight measurement to be determined as we have more equations than variables.  However, not all of the $25$ harmonics are independent;  numerical simulations with random chevrons show that of these $25$ only $19$ are linearly independent, and thus the general problem of depolarizing chevrons cannot be solved with our technique.

However, for chevrons close to NCS form, one can model the most important depolarizing effects.  Consider that the incident beam strikes the chevron's mirrors at grazing incidence, and so the large beam profile averages the Mueller matrices over its area.  To treat this possibility, it is possible to relax the condition that $N^2 + C^2 + S^2 = 1$, corresponding to many different Mueller matrices being averaged over by the beam.  Such a straight fit only involves two extra degrees of freedom.

\subsection{Numerical Verification of Decomposition}

\begin{figure}
    \centering
    \includegraphics[width=0.5\linewidth]{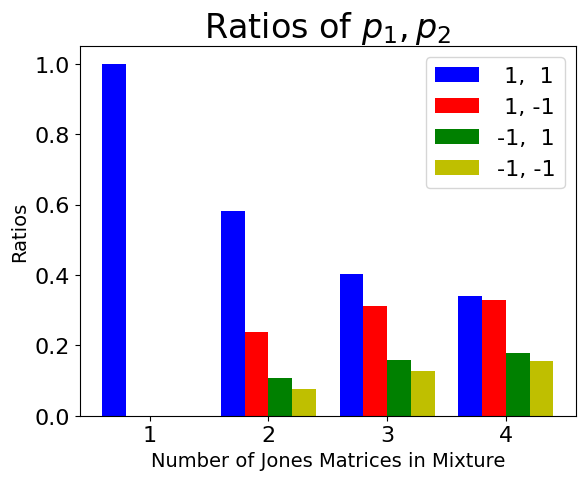}
    \caption{Ratios of Mueller matrices that factor into forms of $p_1, p_2$.  A nondepolarizing matrix always has $p_1 = p_2 = 1$, but mixtures in general may be any one of all four.}
    \label{fig:ratios}
\end{figure}

Random Mueller matrices were generated by creating convex combinations of valid nondepolarizing matrices.  The generalized decomposition was performed and it was verified that all four combinations of $p_1, p_2$ were accessed in the decompositions: see Fig.~\ref{fig:ratios}.  It was also observed that the magnitude of $A$ and $B$ could at times exceed $1$, giving a nonphysical Mueller matrix.  This situation implies that the Mueller matrix is far from NCS form, and a different parameterization would be more appropriate.  

\section{Experimental verification via measurement of a mirror}
\label{sec:data}

The rotating compensator ellipsometer that NIST has developed was first tested using $266$ nm light by frequency-doubling a commercial CW 532 nm laser; this allowed for testing the performance of the system's optics outside of vacuum, and evaluating the performance of the rotating chevrons as compensators by comparing with commercial ellipsometers. Measurements were taken in both a straight geometry, as well as using a sample of a gold mirror in reflection geometry at an incidence angle of 65°. Further details of the instrument design and measurement apparatus are described in a separate paper \cite{Balakrishnan2026}.

\begin{figure}
\centering
\includegraphics[width=0.45\columnwidth]{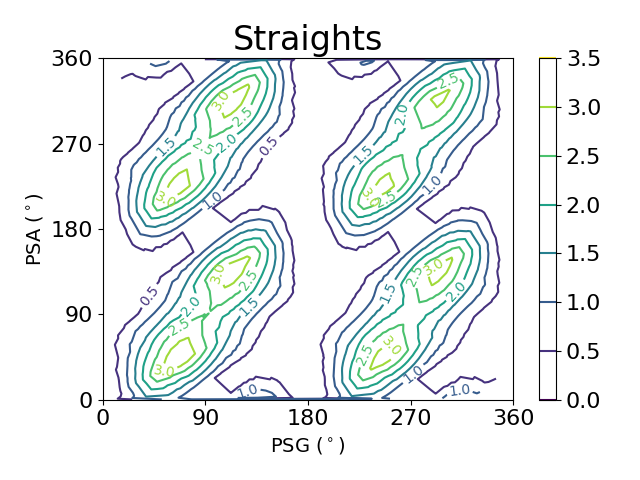}
\includegraphics[width=0.45\columnwidth]{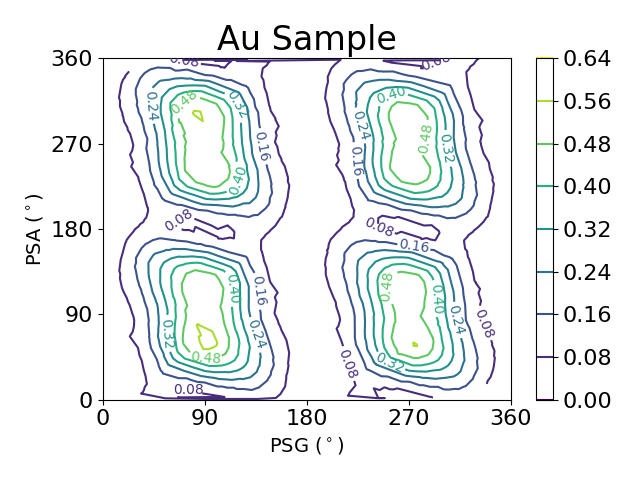}
\caption{Data from a prototype ellipsometer used to determine the ellipsometric parameters N,C, and S of an Au mirror at $\lambda$ = 266 nm. \emph{Left:} Contour plots of measured intensity of light as a function of rotation angles of the PSG and PSA in the straight geometry.  \emph{Right:} Contour plots after reflection off of a thick Au film at an incidence angle of 65° (25° grazing).}
\label{fig:data}
\end{figure}

Fig.~\ref{fig:data} shows the measured signal $S$, the intensity of light normalized to the beam intensity by picking off a portion of the beam; thus, the intensity ratio is unitless. A small asymmetry in the signal can be seen when rotating either the PSG or PSA by 180°. This could be caused either by misalignment of the optical axis and the rotational axis or by occlusion of portions of the beam (mechanical vignetting). The vignetting is accounted for by modulating the occluded beam as a paraboloid restricted to the interval [0,1] that multiplies the overall signal.

\begin{align}
\tilde{S}(\theta_\text{PSA}, \theta_\text{PSG}) &=  \mathcal{O}\left(\theta_\text{PSA}, \theta_\text{PSG}\right) S(\theta_\text{PSA}, \theta_\text{PSG}) \label{eq:straights_modded} \\
\mathcal{O}\left(\theta_\text{PSA}, \theta_\text{PSG}\right) &= \text{min} [1.0, c_\text{PSG}(\theta_\text{PSG} - \theta_{\text{PSG}, 0})^2 + c_\text{PSA}(\theta_\text{PSA} - \theta_{\text{PSA}, 0})^2 + d], ~~~c_\text{PSA}, c_\text{PSG}, d  \geq 0,
\end{align} where $c_\text{PSG}, c_\text{PSA}, \theta_{\text{PSG}, 0}, \theta_{\text{PSA}, 0}, d$ are parameters modeling a paraboloid intended to capture the mechanical vignetting.

The nondepolarizing matrix decomposition was used to find the relevant optical properties of the PSA, PSG, and the NCS values of the optic, as per Eq.~\ref{eq:nondepol}. The straight measurements were fit to Eq.~\ref{eq:straights_modded} using 13 ellipsometric parameters + 5 vignetting parameters.  The measured $N,C,S$ values of the gold mirror were found to match with measurements taken by a commercial ellipsometer, within in-sample variations.  The measured values for N,C, and S are reported in Table~\ref{tab:data}.  The straight calibration found the normalized Stokes and Mueller matrices to be
\begin{align}
D_\text{PSA} &= \begin{bmatrix}
    1 & 0.878 & 0 & 0 \\
    0.878 & 1 & 0 & 0 \\
    0 & 0 & 0.377 & -0.294 \\
    0 & 0 & 0.294 & 0.377
\end{bmatrix} \pm \begin{bmatrix}
    0 & 0.001 & 0 & 0 \\
    0.001 & 0 & 0 & 0 \\
    0 & 0 & 0.002 & 0.004 \\
    0 & 0 & 0.004 & 0.002
\end{bmatrix}, \nonumber \\
D_\text{PSG} &= \begin{bmatrix}
    1 & 0.880 & 0 & 0 \\
    0.880 & 1 & 0 & 0 \\
    0 & 0 & 0.306 & -0.363 \\
    0 & 0 & 0.363 & 0.306
\end{bmatrix} \pm \begin{bmatrix}
    0 & 0.001 & 0 & 0 \\
    0.001 & 0 & 0 & 0 \\
    0 & 0 & 0.001 & 0.001 \\
    0 & 0 & 0.001 & 0.001
\end{bmatrix}, \nonumber \\
P' &= \begin{bmatrix}
    1 & 0.527 & 0.020 & -0.022
\end{bmatrix} \pm \begin{bmatrix}
    0 & 0.001 & 0.001 & 0.008
\end{bmatrix},\nonumber \\ 
Q' &= \begin{bmatrix}
    1 \\
    -0.999 \\
    -0.007 \\
    -0.015
\end{bmatrix} \pm \begin{bmatrix}
    0 \\
    0.001 \\
    0.001 \\
    0.001
\end{bmatrix},\nonumber \\ 
T' &= \begin{bmatrix}
    1.000 & 0 & 0 & 0.008 \\
    0 & 1.000 & -0.021 & 0 \\
    0 & 0.021 & 1.000 & 0 \\
    0.008 & 0 & 0 & 1.000
\end{bmatrix} \pm \begin{bmatrix}
    0.001 & 0 & 0 & 0.003 \\
    0 & 0.001 & 0.002 & 0 \\
    0 & 0.002 & 0.001 & 0 \\
    0.003 & 0 & 0 & 0.001
\end{bmatrix}
\label{eq:fit_straight_params}.
\end{align}  All uncertainties are reported as estimates of the standard uncertainty.

\begin{table}[H]
    \caption{Comparison of spectroscopic ellipsometry (SE) results to a commercial instrument. Standard deviations for the commercial SE represent the standard deviation across measurements at 5 different spots on the sample combined with the uncertainty of the device itself.}
    \label{tab:data}
    \centering
    \vspace{0.5cm}
    \begin{tabular}{c|c|c}
         & NIST SE & Commercial SE  \\
         \hline
         N & $0.534 \pm 0.001$ & ~$0.539 \pm 0.004$ \\
         C & $-0.120 \pm 0.002$ & $-0.123 \pm 0.007$ \\
         S & $0.837 \pm 0.001$ & $0.833 \pm 0.003$
    \end{tabular}
\end{table}

The fits reported in Table~\ref{tab:data} and in Eq.~\ref{eq:fit_straight_params} employed an uncertainty analysis using the techniques similar to those in Ref.~\cite{KOH:2001}: by modeling the discrepancy with Gaussian process regression, it is possible to place uncertainty bounds even without a formula modeling the systematic error.  The details of this technique will be reported in upcoming work.  If the systematic error was neglected, the reported uncertainty would be an order of magnitude too small.  We have restricted the uncertainties to be at minimum $0.001$.

It should be noted that while the chevrons of the PSA and PSG were aligned to within $0.1^\circ$, the fit found a relative difference of $0.6^\circ$.  This is consistent with the misalignment of the mirrors within the chevrons inducing roll in the beam, and would not have been identified if the Mueller matrices of the chevrons were forced to be of $M_\text{diag}(\tau, N, C, S)$ form.

\section{Conclusion}
In this work, we have derived a Mueller matrix decomposition that separates the diagonal NCS form from rotations and circular diattenuation.  The decomposition is invaluable for fitting ellipsometers that involve rotating optics with significant diattenuation and retardance -- even without complete calibration, it is possible to extract meaningful information about samples.  The technique has been applied to measurements performed with NIST's EUV ellipsometer, where the chevrons possess both unknown diattenuation and retardance.  The depolarizing case is also derived, and its applicability in more general ellipsometers has been discussed.  We expect our technique to be employed in future work with our EUV ellipsometer, and in general, any ellipsometer in which standard calibration strategies are difficult or impossible.

\acknowledgments 
 
This work was performed with funding from the CHIPS Metrology Program, part of CHIPS for America, National Institute of Standards and Technology, U.S. Department of Commerce.

\section*{DATA AVAILABILITY STATEMENT}
Data and code used to calculate quantities and generate figures in this manuscript are available upon request.

\section*{CONFLICTS OF INTEREST}
The authors declare no conflicts of interest.

\section*{COPYRIGHT DISCLAIMER}
Official contribution of the National Institute of Standards and Technology; not subject to copyright in the United States.

\bibliography{Refs.bib} 
\bibliographystyle{spiebib} 

\appendix

\section{Proof of Nondepolarizing Decomposition}
\label{app:nondepolarizing}

For nondepolarizing optics, it is more convenient to work with the Jones formalism.  In the Jones formalism, the matrices $J$ are $2 \times 2$ and complex, and are related to the Mueller matrices $M$ by the formula
\begin{align}
M_\text{nondepol}(J) = A [J \otimes J^*] A^{-1}, 
A = \begin{bmatrix}
    1 & 0 & 0 & 1 \\
    1 & 0 & 0 & -1 \\
    0 & 1 & 1 & 0 \\
    0 & i & -i & 0
\end{bmatrix}.
\end{align}
A diagonal Jones matrix (one that preserves $s$ and $p$ polarization) has a Mueller matrix 
\begin{align}
M_\text{diag} (\tau, N, C, S) = \tau \begin{bmatrix}
1 & N & 0 & 0 \\
N & 1 & 0 & 0 \\
0 & 0 & C & S \\
0 & 0 & -S & C
\end{bmatrix},
\end{align} with $N^2 + C^2 + S^2 = 1$;  in the Jones formalism,
\begin{align}
J_\text{diag} (r_s, r_p) = \begin{bmatrix}
r_s & 0 \\
0 & r_p
\end{bmatrix}.
\end{align}
In the Jones formalism, rotations read $R(\theta) = e^{i\theta \sigma_y}$, with $\sigma_i$ denoting the $i$th Pauli matrix
\begin{align}
\sigma_x = \begin{bmatrix}
0 & 1 \\
1 & 0
\end{bmatrix},~
\sigma_y = \begin{bmatrix}
0 & -i \\
i & 0
\end{bmatrix},~
\sigma_z = \begin{bmatrix}
1 & 0 \\
0 & -1
\end{bmatrix}
\end{align}
and note the only matrices that commute with all rotations take the form 
\begin{align}
T = r + s\sigma_y,
\end{align}  with $r, s$ complex numbers. \emph{Almost} every $2 \times 2$ complex matrix may be expressed as 
\begin{align}
J &= a (I + b \sigma_y) (I + c \sigma_z) (I + d \sigma_y)
\end{align} with $a, b, c, d$ complex \footnote{Exceptions form a set of measure 0.}.  This is akin to an Euler angle rotation for complex $2 \times 2$ matrices.  This claim can be proven by expanding the product and showing there is a value for $a, b, c, d$ that yields (almost) any matrix:
\begin{align}
J &= a (1 + bd) + ac (1-bd) \sigma_z + a(b+d) \sigma_y + iac (b-d) \sigma_x \\
&= wI + z\sigma_z + y \sigma_y + x \sigma_x,
\end{align} and $a,b,c,d$ can be expressed in terms of $w, x, y, z$.  Here $I$ is the identity matrix.  There are two solutions, one of which is:
\begin{align}
a &= \dfrac{1}{2} \left(w - \dfrac{\sqrt{(w^2-y^2)z^2(x^2 + z^2)}}{x^2+z^2} \right), \\
b &= \dfrac{-i xyz + wz^2 + \sqrt{(w^2 - y^2) z^2 (x^2 + z^2)}}{z(-i wx + yz)}, \\
c &= - \dfrac{z(x^2 + z^2)}{\sqrt{(w^2 - y^2)z^2(x^2 + z^2)}}, \\
d &= \dfrac{i xyz + wz^2 + \sqrt{(w^2 - y^2) z^2 (x^2 + z^2)}}{z(i wx + yz)},
\end{align}  The other solution can be found employing the identity
\begin{align}
-\sigma_z &= \sigma_y \sigma_z \sigma_y \\
a (I + b \sigma_y) (I + c \sigma_z) (I + d \sigma_y) &= a (b I + \sigma_y) (I - c \sigma_z) (d I + \sigma_y) \\
&= abd (I + \dfrac{1}{b}\sigma_y) (I - c \sigma_z) (I + \dfrac{1}{d}\sigma_y),
\end{align} that is, the transform 
\begin{align}
a \rightarrow abd,~b \rightarrow \frac{1}{b}, c \rightarrow -c, d \rightarrow \dfrac{1}{d}
\end{align} yields the second solution.  The sole caveat is that there may be no solution if any of the quantities $x^2 + z^2, w^2 - y^2, -iwx + yz, iwx + yz$ are $0$, as the denominator for several of the expressions tends to $0$ ($z \rightarrow 0$ is not an issue).  For these exceptional cases, the matrix decomposition fails \footnote{This is not an issue.  An alternate decomposition, employing a $D$ matrix of a different nature, would suffice.}.

We then turn the Jones matrices to Mueller matrices.  If $z = \alpha - i \beta$, then turning $e^{z\sigma_y}$ into a Mueller matrix gives \footnote{We have turned $r(I + b\sigma_y)$ into $c e^{z \sigma_y}$.  This transformation is valid \emph{almost} everywhere: it only fails at $ b = \pm 1$.} gives
\begin{align}
T(\alpha, \beta) &= \begin{bmatrix}
\cosh 2\alpha & 0 & 0 & \sinh 2\alpha \\
0 & \cos 2\beta & -\sin 2\beta & 0 \\
0 & \sin 2\beta & \cos 2\beta & 0 \\
\sinh 2\alpha & 0 & 0 & \cosh 2\alpha
\end{bmatrix}
\end{align}   Meanwhile, the diagonal matrix $aI + b\sigma_z$ corresponds to a Mueller matrix of NCS form.  Thus, for a pure matrix $M_\text{nondepol}$, there is a decomposition 
\begin{align}
    M_\text{nondepol} = T(\alpha_1, \beta_1) M_\text{diag}(\tau, N, C, S) T(\alpha_2, \beta_2),
\end{align} with the $T$ matrices satisfying
\begin{align}
[T(\alpha, \beta), R(\theta)] &= 0 \text{~for all~} \theta\\
T(\alpha_1, \beta_1) T(\alpha_2, \beta_2) &= T(\alpha_1 + \alpha_2, \beta_1 + \beta_2).
\end{align}  These $T$ matrices can be understood as the boost and rotation aspect of the Lorentz group \cite{Devlaminck:08, Simon:10}.  While the $\beta$ quantity in $T$ is a rotation, the $\alpha$ quantity corresponds to circular diattenuation.  Because $\cosh$ is always greater than $1$, this operator $T$ by itself does not correspond to a physical process.  One may normalize the matrix by $e^{-2|\alpha|}$ if one wishes to restrict to physically realizable Mueller matrices.

\section{Proof of Depolarizing Decomposition}
\label{app:depolarizing}

One must work with the full Mueller matrix formalism when investigating the depolarizing matrices.  For notational convenience, let us exchange the second and fourth rows and columns of the Mueller matrix.  Then, the rotation matrix reads 
\begin{align}
R(\theta) = \begin{bmatrix}
    1 & 0 & 0 & 0 \\
    0 & 1 & 0 & 0 \\
    0 & 0 & \cos 2\theta & \sin 2\theta \\
    0 & 0 & -\sin 2\theta & \cos 2\theta
\end{bmatrix}.
\end{align}  The only matrices that commute with all rotations take the form 
\begin{align}
L(U, \theta') = \begin{bmatrix}
    u_{11} & u_{12} & 0 & 0 \\
    u_{21} & u_{22} & 0 & 0 \\
    0 & 0 & \cos 2\theta' & \sin 2\theta' \\
    0 & 0 & -\sin 2\theta' & \cos 2\theta'
\end{bmatrix},
\end{align} up to a multiplicative constant.  Once again, we wish to factor a Mueller matrix $M$ as 
\begin{align}
M = L(U, \theta_1) M_\text{special}^{p_1, p_2} (\tau, N, C, S, A, B) L(V, \theta_2),
\end{align} where the exact form of $M_\text{special}$ is yet to be determined.  However, we know that diagonal matrices of NCS form must be included in $M_\text{special}$, and that there are $6$ degrees of freedom, as the Mueller matrix has $16$ and each $L$ has $5$. 
The form of $M_\text{special}$ is derived by construction, and relies on the singular value decomposition and the QR decomposition.  Recall for any real matrix $M$, the singular value decomposition reads
\begin{align}
M = U \Sigma V^T,
\end{align} where $U, V$ are orthogonal and $\Sigma$ is the nonnegative diagonal matrix of singular values.  A special occurrence happens when the matrices are $2 \times 2$.  In this case, orthogonal matrices are either rotations $R(\theta)$ or Householder reflections $R(\theta)\sigma_z$.  Thus, 
\begin{align}
M &= R(\theta_1) \Sigma R(\theta_2), 
\end{align}  where the `singular values' $\Sigma$ do not need to be positive.  This formula is derived by realizing that if either $U$ or $V$ are Householder reflections, the $\sigma_z$ matrix multiplies $\Sigma$ and changes the sign of one of the singular values to be negative.

The QR decomposition (and the QL decomposition) factor a matrix into an orthogonal times a triangular matrix.  A further decomposition allows one to decompose the triangular matrix as a Heisenberg matrix times a diagonal matrix:

\begin{align}
N = Q R = Q \begin{bmatrix}
a & b \\
0 & c
\end{bmatrix} = Q  \begin{bmatrix}
1 & b/c \\
0 & 1
\end{bmatrix} \begin{bmatrix}
a & 0 \\
0 & c
\end{bmatrix},
\end{align} so long as $c \neq 0$ \footnote{The case where $c = 0$ will be left to future work.  Again, however, this set of problematic matrices forms a set of measure 0.}.  The Heisenberg matrix always has an inverse:
\begin{align}
\begin{bmatrix}
1 & \alpha \\
0 & 1
\end{bmatrix}^{-1} = \begin{bmatrix}
1 & -\alpha \\
0 & 1
\end{bmatrix}
\end{align}
Now consider a real $4 \times 4$ matrix, but block the elements into four matrices of size $2 \times 2$.  This matrix $M$ is the depolarizing Mueller matrix with the second and fourth rows and columns exchanged.
\begin{align}
M = \begin{bmatrix}
A & B \\
C & D
\end{bmatrix}.
\end{align}  Perform a generalized singular value decomposition on $D = R(\theta_1) \Sigma R(\theta_2)$, allowing the entries of $\Sigma$ to be negative, and enforce the second singular value of $\Sigma$ to have larger magnitude than the first and be positive.  If the largest magnitude singular value is negative, multiply both $\Sigma$ and $R(\theta_2)$ by $-1$.  $-R(\theta_2) = R(\theta_2 + \frac{\pi}{2})$ still remains a rotation and $\Sigma$ now has a positive largest singular value.

\begin{align}
M = \begin{bmatrix}
I & 0 \\
0 & R(\theta_1)
\end{bmatrix} \begin{bmatrix}
A & B R(-\theta_2) \\
R(-\theta_1) C & \Sigma
\end{bmatrix} \begin{bmatrix}
I & 0 \\
0 & R(\theta_2)
\end{bmatrix},
\end{align}
and now we employ a QR or QL decomposition on $R(-\theta_1)C$ and $B R(-\theta_2)$.  Let
\begin{align}
B R(-\theta_2) = Q_3 H_3 D_3 \\
R(-\theta_1) C = D_4 H_4 Q_4,
\end{align} with $H$ a Heisenberg matrix (either lower triangular or upper triangular) and $D$ a diagonal matrix.
\begin{align}
M &= \begin{bmatrix}
Q_3 H_3 \sigma_x & 0 \\
0 & R(\theta_1)
\end{bmatrix} \begin{bmatrix}
A' & \sigma_x D_3 \\
D_4 \sigma_x & \Sigma
\end{bmatrix} \begin{bmatrix}
\sigma_x H_4 Q_4 & 0 \\
0 & R(\theta_2)
\end{bmatrix},\\
A' &= \sigma_x H_3^{-1} Q_3^T A Q_4^T H_4^{-1} \sigma_x.
\end{align}  Let us define
\begin{align}
A' = \begin{bmatrix}
    a_{11} & a_{12} \\
    a_{21} & a_{22}
\end{bmatrix}, ~\sigma_x D_3 = \begin{bmatrix}
    0 & \tau N_R \\
    -\tau S_R & 0
\end{bmatrix}, ~D_4 \sigma_x = \begin{bmatrix}
    0 & \tau S_L \\
    \tau N_L & 0
\end{bmatrix}, ~\Sigma = \begin{bmatrix}
    \tau C & 0 \\
    0 & \tau
\end{bmatrix}, \\
U = Q_3 H_3 \sigma_x = \begin{bmatrix}
    u_{11} & u_{12} \\
    u_{21} & u_{22}
\end{bmatrix},~
V = \sigma_x H_4 Q_4 = \begin{bmatrix}
    v_{11} & v_{12} \\
    v_{21} & v_{22}
\end{bmatrix}
\end{align}

\begin{align}
M = \begin{bmatrix}
u_{11} & u_{12} & 0 & 0 \\
u_{21} & u_{22} & 0 & 0 \\
0 & 0 & \cos 2\theta_1 & \sin 2\theta_1 \\
0 & 0 & -\sin 2\theta_1 & \cos 2\theta_1 
\end{bmatrix} \begin{bmatrix}
    a_{11} & a_{12} & 0 & \tau N_R \\
    a_{21} & a_{22} & -\tau S_R & 0 \\
    0 & \tau S_L & \tau C & 0 \\
    \tau N_L & 0 & 0 & \tau
\end{bmatrix} \begin{bmatrix}
v_{11} & v_{12} & 0 & 0 \\
v_{21} & v_{22} & 0 & 0 \\
0 & 0 & \cos 2\theta_2 & \sin 2\theta_2 \\
0 & 0 & -\sin 2\theta_2 & \cos 2\theta_2 
\end{bmatrix}.
\end{align}  As a final step, we shall apply a diagonal scale transform on the upper left block to turn the middle matrix into $M_\text{special}$ form:
\begin{align}
M_\text{special}^{(p_1, p_2)}(\tau, N, C, S, A, B) = \tau
\begin{bmatrix}
    1 & B & 0  & p_1 N \\
    A & C & -p_2 S  & 0 \\
    0 & S & C & 0 \\
    N & 0 & 0 & 1 
\end{bmatrix},~~p_1 = \pm 1,~ p_2 = \pm 1.
\end{align}

\begin{align}
\begin{bmatrix}
r_1 & 0 & 0 & 0 \\
0 & s_1 & 0 & 0 \\
0 & 0 & 1 & 0 \\
0 & 0 & 0 & 1
\end{bmatrix}
\begin{bmatrix}
    a_{11} & a_{12} & 0 & \tau N_R \\
    a_{21} & a_{22} & -\tau S_R & 0 \\
    0 & \tau S_L & \tau C & 0 \\
    \tau N_L & 0 & 0 & \tau
\end{bmatrix}
\begin{bmatrix}
r_2 & 0 & 0 & 0 \\
0 & s_2 & 0 & 0 \\
0 & 0 & 1 & 0 \\
0 & 0 & 0 & 1
\end{bmatrix} = \begin{bmatrix}
    r_1 r_2 a_{11} & s_2 r_1 a_{12} & 0 & \tau r_1 N_R \\
    s_1 r_2 a_{21} & s_1 s_2 a_{22} & -\tau s_1 S_R & 0 \\
    0 & \tau s_2 S_L & \tau C & 0 \\
    \tau r_2 N_L & 0 & 0 & \tau
\end{bmatrix}.
\end{align}  To match the form of $M_\text{special}$, we require
\begin{align}
r_1 r_2 a_{11} &= \tau , ~~r_2 N_L = p_1 r_1 N_R \\
s_1 s_2 a_{22} &= \tau C, ~~s_1 S_R = p_2 s_2 S_L.
\end{align}
In order for these two systems to have real solutions $r_1, r_2, s_1, s_2$, the products $p_1 a_{11} \tau N_L N_R$ and $p_2 a_{22} \tau C S_L S_R$ must be nonnegative.  If either product is negative, there will exist a solution $r_1, r_2, s_1, s_2$ but some variables will be imaginary, which is disallowed in a Mueller matrix.  This is the reason for the $p_1, p_2$ variables that can take on the values $\pm 1$: they enforce the sign condition so that there exists real solutions for $r_1, r_2, s_1, s_2$.  The physical implications of such a matrix are unclear.  The situation where either product is $0$ is left to future work.

The solution to the system of equations reads
\begin{align}
p_1 &= \text{sgn}[\tau a_{11} N_L N_R] \\
p_2 &= \text{sgn}[\tau C a_{22} S_L S_R] \\
r_1 &= \sqrt{\dfrac{\tau N_L}{p_1 a_{11} N_R}},~~r_2 = \text{sgn}\left[\dfrac{\tau}{a_{11}} \right] \sqrt{\dfrac{p_1 \tau N_R}{ a_{11} N_L}} \\
s_1 &=\sqrt{\dfrac{p_2 \tau C S_L}{ a_{22} S_R}},~~s_2 = \text{sgn}\left[\dfrac{\tau C}{a_{22}}  \right]  \sqrt{\dfrac{\tau C S_R}{p_2 a_{22} S_L}}.
\end{align}

Numerical simulations show that there are valid Mueller matrices that, when decomposed, correspond to an $M_\text{special}$ with $(p_1, p_2) \neq (1, 1)$: see Fig.~\ref{fig:ratios}.

The final factorization reads
\begin{align}
M = \begin{bmatrix}
u_{11}/r_1 & u_{12}/s_1 & 0 & 0 \\
u_{21}/r_1 & u_{22}/s_1 & 0 & 0 \\
0 & 0 & \cos 2\theta_1 & \sin 2\theta_1 \\
0 & 0 & -\sin 2\theta_1 & \cos 2\theta_1 
\end{bmatrix} M_\text{special}^{(p_1, p_2)}(\tau, N, C, S, A, B) \begin{bmatrix}
v_{11}/r_2 & v_{12}/r_2 & 0 & 0 \\
v_{21}/s_2 & v_{22}/s_2 & 0 & 0 \\
0 & 0 & \cos 2\theta_2 & \sin 2\theta_2 \\
0 & 0 & -\sin 2\theta_2 & \cos 2\theta_2 
\end{bmatrix}.
\end{align}
Exchanging the second and fourth rows and columns to revert back to the original form of the Mueller matrix completes the decomposition.

\end{document}